\documentclass{aa}
\usepackage[varg]{txfonts}
\usepackage[caption = false]{subfig}
\usepackage{tabularx}
\usepackage{graphicx}
\usepackage{amsmath}
\usepackage{amssymb}
\usepackage{color}
\usepackage[colorinlistoftodos]{todonotes}
\usepackage{bm}
\usepackage{float}
\usepackage[colorinlistoftodos]{todonotes}

\usepackage{natbib,twoopt}
\usepackage[breaklinks=true]{hyperref} 
\bibpunct{(}{)}{;}{a}{}{,}             
\makeatletter
  \newcommandtwoopt{\citeads}[3][][]{\href{http://adsabs.harvard.edu/abs/#3}%
    {\def\hyper@linkstart##1##2{}%
     \let\hyper@linkend\@empty\citealp[#1][#2]{#3}}}
  \newcommandtwoopt{\citepads}[3][][]{\href{http://adsabs.harvard.edu/abs/#3}%
    {\def\hyper@linkstart##1##2{}%
     \let\hyper@linkend\@empty\citep[#1][#2]{#3}}}
  \newcommandtwoopt{\citetads}[3][][]{\href{http://adsabs.harvard.edu/abs/#3}%
    {\def\hyper@linkstart##1##2{}%
     \let\hyper@linkend\@empty\citet[#1][#2]{#3}}}
  \newcommandtwoopt{\citeyearads}[3][][]%
    {\href{http://adsabs.harvard.edu/abs/#3}
    {\def\hyper@linkstart##1##2{}%
     \let\hyper@linkend\@empty\citeyear[#1][#2]{#3}}}
\makeatother

\hypersetup{draft}

\begin{document}

\title{Statistical model for filamentary structures of molecular clouds}
\subtitle{The modified multiplicative random cascade model and its multifractal nature}

\author{J.-F.~Robitaille\inst{\ref{inst1}}, A.~Abdeldayem\inst{\ref{inst1}}, I.~Joncour\inst{\ref{inst1}}, E.~Moraux\inst{\ref{inst1}}, F.~Motte\inst{\ref{inst1}}, P.~Lesaffre\inst{\ref{inst2}}, A.~Khalil\inst{\ref{inst3}}}
\authorrunning{J.-F.~Robitaille et al.}

\institute{Univ. Grenoble Alpes, CNRS, IPAG, 38000 Grenoble, France,~\email{jean-francois.robitaille@univ-grenoble-alpes.fr}\label{inst1}
                \and
                Laboratoire de Physique de l’Ecole normale supérieure, ENS, Université PSL, CNRS, Sorbonne Université, Université Paris- Diderot, Sorbonne Paris Cité, Paris, France\label{inst2}
                \and
                CompuMAINE Laboratory, Department of Mathematics and Statistics, University of Maine, Orono, ME 04469, USA\label{inst3}} 

\date{Accepted: 30/06/2020}

\abstract{We propose a new statistical model that can reproduce the hierarchical nature of the ubiquitous filamentary structures of molecular clouds. This model is based on the multiplicative random cascade, which is designed to replicate the multifractal nature of intermittency in developed turbulence. We present a modified version of the multiplicative process where the spatial fluctuations as a function of scales are produced with the wavelet transforms of a fractional Brownian motion realisation. This simple approach produces naturally a log-normal distribution function and hierarchical coherent structures. Despite the highly contrasted aspect of these coherent structures against a smoother background, their Fourier power spectrum can be fitted by a single power law. As reported in previous works using the multiscale non-Gaussian segmentation (MnGSeg) technique, it is proven that the fit of a single power law reflects the inability of the Fourier power spectrum to detect the progressive non-Gaussian contributions that are at the origin of these structures across the inertial range of the power spectrum. The mutifractal nature of these coherent structures is discussed, and an extension of the MnGSeg technique is proposed to calculate the multifractal spectrum that is associated with them. Using directional wavelets, we show that filamentary structures can easily be produced without changing the general shape of the power spectrum. The cumulative effect of random multiplicative sequences succeeds in producing the general aspect of filamentary structures similar to those associated with star-forming regions. The filamentary structures are formed through the product of a large number of random-phase linear waves at different spatial wavelengths. Dynamically, this effect might be associated with the collection of compressive processes that occur in the interstellar medium.}

\maketitle 

\section{Introduction}

The origin of the scale-free nature of the interstellar medium (ISM) has been debated for several decades. One of the first scale-free signature, the Fourier power spectrum of Galactic HI emission, was attributed to the turbulent nature of the ISM \citepads{1993MNRAS.262..327G}. Turbulence naturally imposes scale-free hierarchical structures to the ISM, where large eddies are extended and subdivided into smaller ones to which they transfer a fraction of their energy up to the dissipation scale \citepads{1922Richardson, 2019PhRvE..99d2113M}. However, comparisons with statistical models of cloud structure proved that scale-free density fluctuations do not succeed in reproducing the ubiquitous filamentary structures seen in the ISM \citepads{2001ApJ...548..749E, 2007A&A...469..595M}. More recently, some studies on thermal dust emission observed by \emph{Herschel} raised the question why contrasted filamentary structures, which appear to have a characteristic width \citepads{2011A&A...529L...6A, 2019A&A...621A..42A}, do not produce any break or kink in the Fourier power spectrum \citepads{2017MNRAS.466.2529P}. \citetads{2014MNRAS.440.2726R, 2019A&A...628A..33R} showed using the multiscale non-Gaussian segmentation (MnGSeg) technique\footnote{Codes and tutorials are available at \url{https://github.com/jfrob27/pywavan}} that filamentary structures are in fact dominating the scale-free Fourier power spectrum. This result demonstrated that dense coherent structures of the ISM also have a hierarchical geometry (most likely multifractal and intimately linked to the formation of stars) and that this geometry is intertwined with the monofractal and diffuse component of molecular clouds. This finding challenges the recent results of \citetads{2019A&A...626A..76R}, who suggested that only filamentary structures with a sufficiently high area filling factor and/or high column density contrasts have an effect on the scale-free power spectrum of dust-continuum images.

Moreover, \citetads{2018MNRAS.481..509E} recently emphasised that the monofractal approach to characterising ISM structures underlies a certain degree of degeneracy as the statistical description of the regions is based on a single parameter, the fractal dimension. They proposed to analyse the multifractal properties of ISM structures in Galactic dust-continuum maps using the box-counting approach. The multifractal mathematical framework, which is now applied in many fields, from financial markets to medical imagery analysis, can detect complex local structures and can describe local singularities. Multifractal analysis is also intimately linked to the study of fully developed turbulence \citepads{1995turb.book.....F}, where instead of trying to capture the turbulence intermittency from the final state of the dissipation scale, the multifractal model suggests that the overall flow might be described as a superposition of coexisting structures at all scales that are correlated with each other by a cascading intensity \citepads{2019PhRvE..99d2113M}.

The multifractal analysis of ISM structures was also performed by \citetads{2006ApJS..165..512K} on Galactic HI-line emission using a wavelet-based formalism. In contrast with the dust continuum, HI maps revealed a monofractal signature in their study. The monofractal signature of the Galactic HI could be attributed on the one hand to the lower resolution of radio observations, and on the other hand, to the more diffuse nature of the HI emission, which is less inclined to produce local singularities than the dust far-infrared emission.

In this paper, we propose a new statistical model to produce synthetic coherent filamentary structures based on the multiplicative cascade process. This new model reproduces some aspects of the probability distribution function (PDF) of young star-forming regions and the scale-free Fourier power spectrum despite the presence of highly contrasted coherent filamentary structures. Moreover, we propose a multifractal analysis framework based on the wavelet formalism of the MnGSeg technique. As discussed in \S\ref{sec:Multifractal}, we strongly suggest using the wavelet approach for the multifractal analysis of continuous images instead of the box-counting formalism developed for discrete sets of points.

The paper is structured as follows: \S \ref{sec:fractal} presents the fractal models of the ISM and the general physical properties assigned to them; \S \ref{sec:random_cascade} presents the details of the multiplicative random cascade model and two modified models based on the wavelet decomposition of random fractal images; \S \ref{sec:Multifractal} presents the wavelet-based multifractal formalism and an analysis of the synthetic monofractal and multifractal models; \S \ref{sec:hierarchy} discusses the hierarchical nature of the models and compares them with observations; and we finally use this new multifractal framework to produce synthetic ISM structures and lay out prospects for future analyses in \S \ref{sec:conclusion}.

\section{Fractal models}\label{sec:fractal}

\begin{figure*}
\centering
\includegraphics[width=0.90\textwidth]{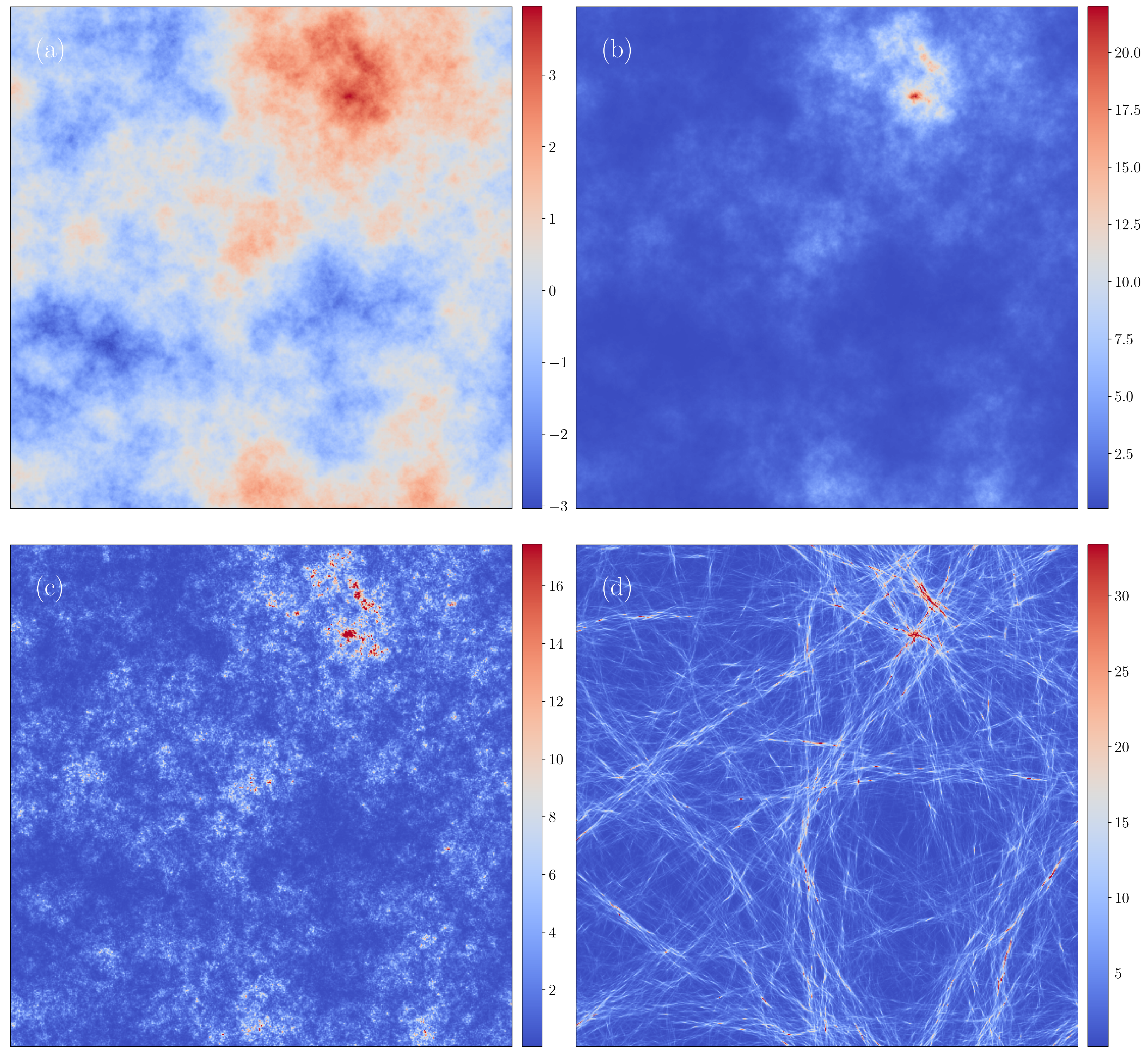}
\caption{Spatial intensity distribution of the fractal models described in Sect. \ref{sec:fractal}. (a) fBm model and (b) exponentiated fBm model, and the random cascade models described in Sect. \ref{sec:random_cascade}. (c) Isotropic and (d) directional random cascade models.}
\label{fig:fractal_models}
\end{figure*}

\begin{figure}
\centering
\includegraphics[width=0.48\textwidth]{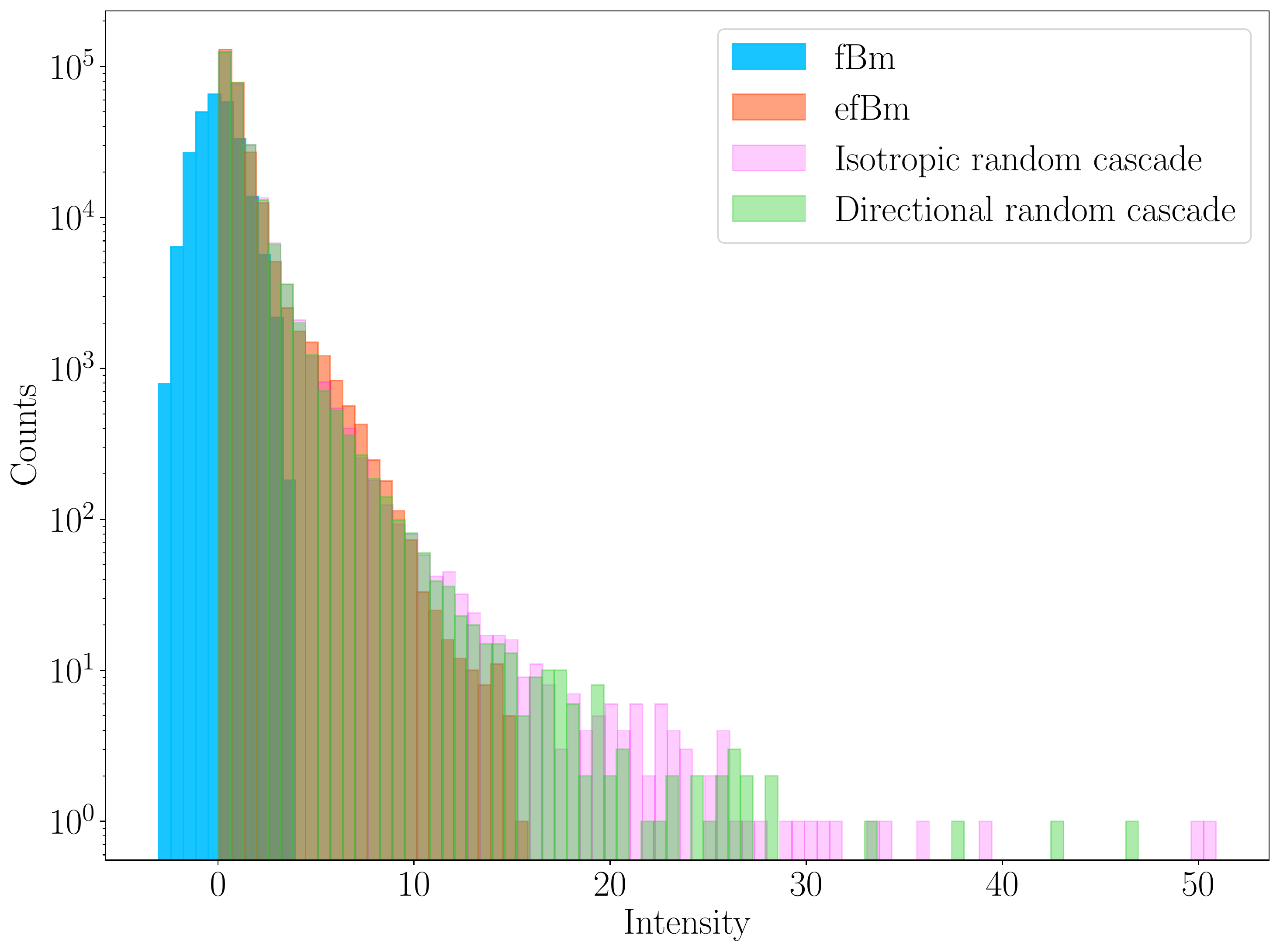}
\caption{Distribution functions for the four models presented in Fig. \ref{fig:fractal_models}. Distribution functions for the efBm, the isotropic, and the directional random cascade are normalised by the mean pixel value of the image.}
\label{fig:pdfs}
\end{figure}

Fractional Brownian motion (fBm) models, as a basic model for spatial intensity fluctuations in the ISM, can reproduce the scale-free nature of structures measured by the Fourier power spectrum. They are produced using the inverse Fourier transform of random phase values in the $u$-$v$ plane multiplied by a power law for the squared modulus of the complex numbers. For a power law of $\sim-3.0$, the distribution function of the fBm model is approximately Gaussian. An fBm model with a power-law index of $-3.1$ is shown in Fig. \ref{fig:fractal_models} (a), and its distribution function is shown in Fig. \ref{fig:pdfs}. The fBm model has a zero mean value and a standard deviation of one.

In order to reproduce the typical log-normal density distribution measured in observations and simulations, \citetads{2001ApJ...548..749E} proposed exponentiating the fBm following the relation

\begin{equation}
\rho_{\rm efbm}(\mathbf{x}) = \exp [\sigma_{\rho}f(\mathbf{x})],
\label{eq:exp_fbm}
\end{equation}

\noindent where $f(\mathbf{x})$ is the fBm image, $\mathbf{x}$ is the position vector, $\sigma_{\rho}$ is the standard deviation attributed to the fBm, and $\rho$ is the modelled gas density. According to the results of \citetads{1997ApJ...474..730P}, $\sigma_{\rho}$ can be related to the Mach number, $\mathcal{M}$, as

\begin{equation}
\sigma_{\rho} = (\ln[1+0.5\mathcal{M}^2])^{0.5}.
\label{eq:zeta}
\end{equation}

\noindent Because $\sigma_{\rho}f(\mathbf{x})=\ln\rho(\mathbf{x})$, the PDF, $P$, can be written as

\begin{equation}
P(\ln\rho) = \frac{1}{(2\pi\sigma_{\rho})^{1/2}}\exp\left[-0.5\left(\frac{\ln\rho-\mu}{\sigma_{\rho}}\right)^2\right],
\label{eq:PDF_dens}
\end{equation}

\noindent where $\mu$ is the arithmetic mean value of $f(\mathbf{x})$. The result of the exponentiated fBm (efBm) is shown in Fig. \ref{fig:fractal_models} (b) for $\mathcal{M}=1.5$. The distribution function of this model is shown in Fig. \ref{fig:pdfs}. As stated by \citetads{2001ApJ...548..749E}, even if this model succeeds in reproducing the log-normal PDF and the single power law of the ISM, the resulting map lacks the  sharp transitions in density that are commonly identified as filaments, shells, and holes in star-forming regions.

\section{Multiplicative random cascade model}\label{sec:random_cascade}

As noted by \citetads{1994ApJ...423..681V}, hierarchical structures are naturally expected to arise in flows in which the density has a log-normal distribution. This emergence of a universal distribution can be associated with the central limit theorem, for which the sum of a large number of random variables converges to a normal distribution. If this assumption is attributed to a random sequence of $\ln\rho$, then the density $\rho$ becomes log-normally distributed. Thus, according to this description, after a finite time, density can be considered the product of a large number of independent random fluctuations $\delta$,

\begin{equation}
\rho(t_n)=\delta_n\delta_{n-1}\textrm{... }\delta_1\delta_0\rho(t_0)
\label{eq:multi_density}
.\end{equation}

\noindent In this paper, we point out the correspondence between Eq. \ref{eq:multi_density} and the multiplicative random cascade model that is designed to reproduce the multifractal nature of intermittency in developed turbulence \citepads{1991JFM...224..429M, 1995turb.book.....F}. In this model, the rate of turbulent energy dissipation $\varepsilon$ is considered first as a nonrandom positive quantity for a cubic volume of side $\ell_0$. This volume is then subdivided into eight equal cubes of side $\ell_1 = \ell_0/2$. For every subdivisions, the dissipation $\varepsilon$ is multiplied by a series of independent random variables $W_n$ that are distributed identically. After $n$ generations, the dissipation value of a cube of side $\ell_n = \ell_02^{-n}$  becomes $\varepsilon_{\ell}=W_nW_{n-1}\textrm{... }W_1W_0\varepsilon$.

The sequence of products by an independent random variable, $\delta$ in the case of density fluctuations, has the effect of producing large deviations, that is, an increasingly non-Gaussian statistics towards smaller scales. In the ISM, this effect is measured notably through anomalous scaling of the high-order structure functions of centroid velocity increments \citepads{1996ApJ...463..623L,2008A&A...481..367H, 2010A&A...512A..81F, 2015MNRAS.446.3777B} (see \S\ref{sec:Multifractal}). It has been observed in the early 1990s that a probabilistic reformulation of turbulence intermittency analysis in terms of wavelet transforms of the velocity field was a powerful alternative to velocity increments \citepads{1993PhRvE..47..875M, 1995PhyA..213..232A, 1995turb.book.....F}. More recently,  \citetads{2014MNRAS.440.2726R, 2019A&A...628A..33R} discovered a similar anomalous scaling signature, generally attributed to turbulence intermittency, in \emph{Herschel} column density maps using PDFs of wavelet coefficients at different spatial scales. 

\subsection{Modified random cascade model}

\begin{figure}
\centering
\includegraphics[width=0.48\textwidth]{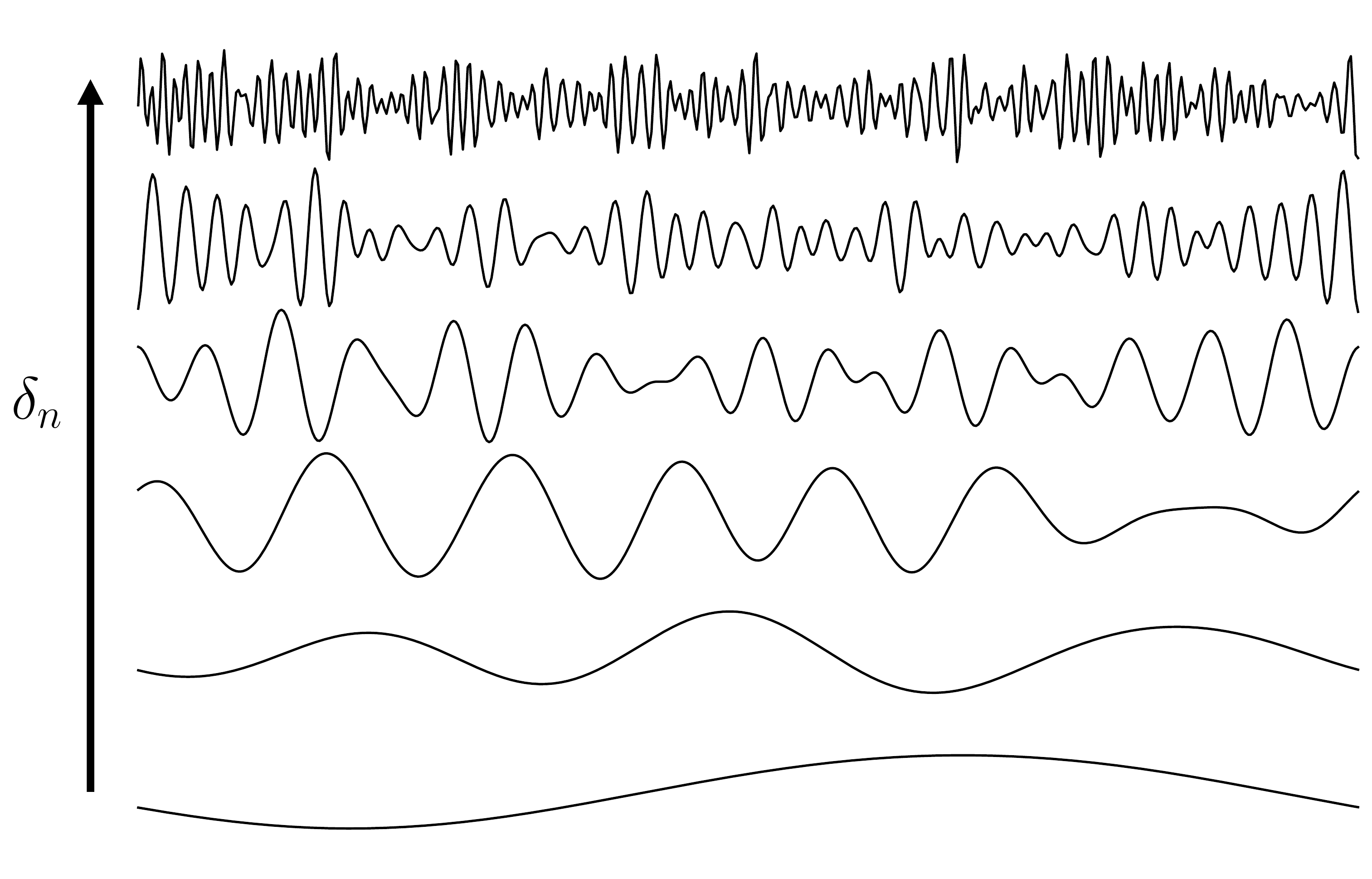}
\caption{Schematic representation of the modified random cascade model, where each level is the normalised spatial fluctuations of an fBm as a function of scales obtained from wavelet transforms.}
\label{fig:cascade_cartoon}
\end{figure}

In this paper, we modify the random cascade model by substituting the layered cubes at different scales with the wavelet transforms of the fBm model presented in section \ref{sec:fractal}. Figure \ref{fig:cascade_cartoon} gives a schematic representation of the independent random variable $\delta$ as multiscale density fluctuations. The wavelet transform of a signal gives the local amount of fluctuations as a function of scales, and, for anisotropic wavelet functions, also as a function of orientations for a 2D signal. Because continuous wavelets are redundant, a signal can be reconstructed from its wavelet coefficients simply by integrating over the scales and the orientations \citepads{2019A&A...628A..33R},

\begin{equation}
f(\mathbf{x}) = C_{\delta} \sum_{\ell} \sum_{j=0}^{N_{\theta}-1} \ell \tilde{f}(\mathbf{x},\ell,\theta_j) + \mu_0,
\label{eq:synthesis}
\end{equation}

\noindent where $C_{\delta}$ is a correction factor, $\mu_0$ is the mean value of the original signal, and $\tilde{f}(\mathbf{x},\ell,\theta_j)$ are the wavelet coefficients calculated from the wavelet transform, defined as

\begin{equation}
\tilde{f}(\mathbf{x} ,\ell, \theta) = \mathcal{F}^{-1}\left\{\hat{f}(\vec{k})\hat{\psi}_{\ell,\theta}^*(\vec{k})\right\}.
\label{eq:wavelet_transform}
\end{equation}

\noindent In equation \ref{eq:wavelet_transform}, $\mathcal{F}^{-1}$ denotes the inverse Fourier transform, and $\hat{f}(\vec{k})$ and $\hat{\psi}_{\ell,\theta}(\vec{k})$ represent the Fourier transform of $f(\mathbf{x})$ and of a daughter version of the mother wavelet, respectively. The latter being dilated to a given scale $\ell$ and rotated by an azimuthal angle $\theta$. In order to recover the multiplicative process described by \citetads{1994ApJ...423..681V} and modelled by the random cascade, Eq. \ref{eq:exp_fbm} can be expressed using the wavelet synthesis relation of Eq. \ref{eq:synthesis},

\begin{equation}
\rho_{\rm mc}(\mathbf{x}) = \exp \left[\sigma \sum_{l,\theta} C_{\ell} \tilde{f}(\mathbf{x},\ell,\theta)\right]
                = \prod_{l,\theta} \exp \left[\sigma C_{\ell} \tilde{f}(\mathbf{x},\ell,\theta)\right],
\label{eq:random_casc_wavelet}
\end{equation}

\noindent where $C_{\ell}$ is a constant depending on scale $\ell$. Equation \ref{eq:random_casc_wavelet} demonstrates that when the fractal image is considered as the summation of many spatial fluctuations, the exponentiated fBm model of equation \ref{eq:exp_fbm} becomes equivalent to the multiplicative process described in equation \ref{eq:multi_density}.

However, compared to the exponentiated fBm, the layered spatial fluctuations resulting from the fBm wavelet decomposition have to be modified. An fBm image, by construction, follows a squared-amplitude power law for its spatial fluctuations. As shown in Fig. \ref{fig:fractal_models} (a), this power law is at the origin of the hierarchal nature of the fractal image, where low spatial frequencies dominate high spatial frequencies $\mathbf{k}$. In order to meet the random cascade model requirements, all spatial frequencies need to be identically distributed. For this reason, the constant $C_{\ell}$ in equation \ref{eq:random_casc_wavelet} becomes a scale-dependent normalisation factor $C_{\ell} = 1 / \sigma_{\ell}$, where $\sigma_{\ell}$ is the standard deviation of wavelet coefficients $\tilde{f}(\mathbf{x},\ell,\theta)$. The construction of a discrete multiplicative cascade, called the `$\mathcal{W}$-cascades', using an orthogonal wavelet basis was proposed by \citetads{1998JMP....39.4142A} to reproduce the multifractal nature of turbulence and of financial time series. More recently, \citetads{2019PhRvE..99d2113M} propose another grid-free model, closer to the one proposed in the present paper, using continuous wavelet transforms. Our model uses 2D continuous wavelet transforms in order to synthesise the multifractal nature of ISM maps.

To calculate the modified random cascade model, we used the Fan wavelet transform \citepads{2005CG.....31..846K, 2019A&A...628A..33R}. The Fan wavelet is a rotated version of the classical Morlet wavelet and is designed to optimally sample all azimuthal directions in Fourier space. The Fourier transform of the Morlet wavelet is defined as

\begin{equation}
\begin{split}
\hat{\psi}_{\ell,\theta}(\vec{k}) & = \ell \cdot \textrm{e}^{-|\vec{k}-\vec{k}_0|^2/2} \\
                                     & = \ell \cdot \textrm{e}^{-[(u-|\vec{k}_0|\cos \theta)^2+(v-|\vec{k}_0|\sin \theta)^2]/2},
\end{split}
\label{eq:Morlet_Fourier}
\end{equation}

\noindent where the vector wavenumber is defined as $\vec{k} = (u, v)$ and the constant $|\vec{k}_0| = \sqrt{u_0^2 + v_0^2}$ is set to $\pi\sqrt{2/\ln2} \approx 5.336$ to ensure that the admissibility condition, $\int^{+\infty}_{-\infty}\psi(x)dx=0$, is almost met \citepads{2005CG.....31..846K}. The wavelet transform is performed following equation \ref{eq:wavelet_transform}. The result of the modified random cascade model as defined in equation \ref{eq:random_casc_wavelet} is shown in Fig. \ref{fig:fractal_models} (c). This model uses the same fBm model as in Fig. \ref{fig:fractal_models} (a). A different standard deviation $\sigma$ is introduced in Eq. \ref{eq:random_casc_wavelet} in order to keep the same value of $\mathcal{M}\simeq1.5$ while taking into account the new coefficient $C_{\ell}$. The distribution function of the modified random cascade model, for the isotropic case, is also shown in Fig. \ref{fig:pdfs}. The next section will explore the anisotropic property of the Morlet wavelet in the construction of the modified random cascade model.

\subsection{Directional modified random cascade model}\label{sec:directional}

Finally, we explored the angular dependence of the modified random cascade model. As for the initial fBm and the efBm, the random cascade also fails to reproduce the typical highly contrasted filamentary structures observed in star-forming regions. Dynamically, the cumulative effect of the random sequence of $\ln \rho$ could be attributed to the accumulation of shock-like structures in the turbulent ISM (velocity shears, supersonic shocks, etc.). These shock fronts are rarely isotropic, and their preferential direction can depend on the large-scale inhomogeneity of the gas distribution before the passage of the shock front. Moreover, according to the \citetads{2015A&A...580A..49I} model, molecular clouds would be preferentially formed in limited regions where the compressional direction of the shock is almost parallel to the local mean magnetic field lines or in regions experiencing a larger number of sequential compressions.

In order to reproduce these anisotropies in the modified random cascade model, we removed the angular dependence in relation \ref{eq:random_casc_wavelet} and performed the multiplicative process for each angle independently before summing the resulting random cascades over all direction. In our model, the angular directions are discretised in 11 directions by the Morlet wavelet transforms in order to sample the Fourier space optimally \footnote{See \citet{2019A&A...628A..33R} section 3.1 and Fig. 2 for a representation of the Morlet wavelet sampling in Fourier space.}. Following this new model, equation \ref{eq:random_casc_wavelet} now becomes

\begin{equation}
\rho_{\rm dmc}(\mathbf{x}) = \frac{1}{N_{\theta}} \sum_{\theta} \left[\prod_{l} \exp \left[\sigma C_{\ell} \tilde{f}(\mathbf{x},\ell,\theta)\right]\right].
\label{eq:ang_random_casc_wavelet}
\end{equation}
                
\noindent The result of the directional modified random cascade model is shown in Fig. \ref{fig:fractal_models} (d). This model succeeds in producing filamentary structures. It uses the same original fBm model as in Fig. \ref{fig:fractal_models} (a) and $\mathcal{M}\simeq1.5$. As shown in Fig. \ref{fig:pdfs}, the intensity distributions of the isotropic and directional random cascade models and of the efBm model are similar. It is worth noting that increasing the number of angular directions does not change the general aspect of the model, nor the straightness of the filamentary structures. Moreover, because the intensity structures in each model are different, this means that the global intensity distribution alone cannot be used to distinguish the different models.

\begin{figure}
\centering
\includegraphics[width=0.48\textwidth]{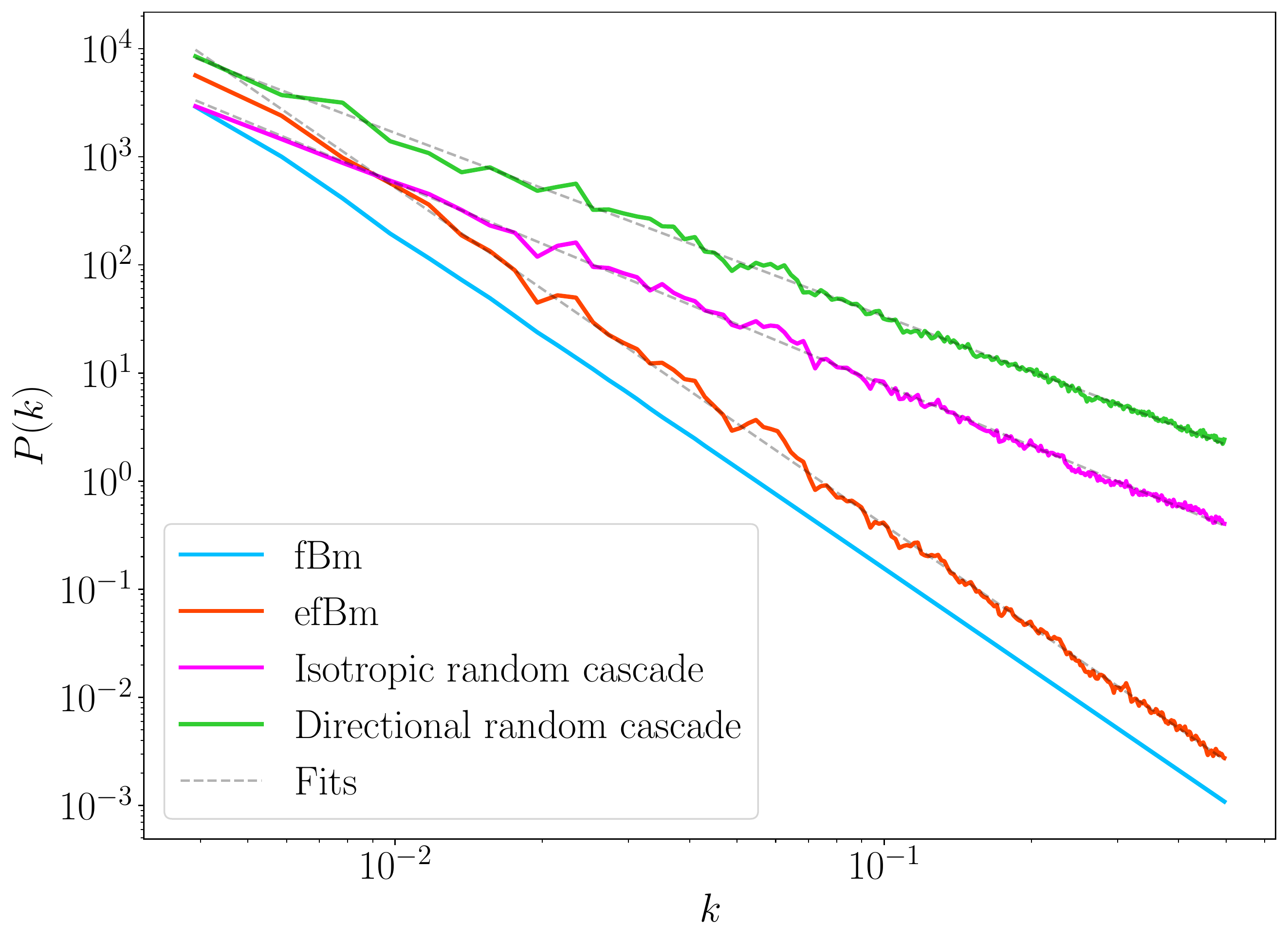}
\caption{Fourier power spectrum for the four models presented in Fig. \ref{fig:fractal_models}. The dashed lines show the fitted power laws. The fitted power indices are listed in Table \ref{tab:pow_spect_fits_models}.}
\label{fig:pow_spec_models}
\end{figure}

The Fourier power spectra for the four models are shown in Fig. \ref{fig:pow_spec_models}. Despite the highly contrasted features and the very different spatial intensity distribution, all power spectra exhibit a single power law.  As shown in Fig. \ref{fig:model_scale_PDFs}, the multiplicative operation in our models produces a progressive non-Gaussian contribution towards the small scales at which the Fourier analysis is not sensitive. Compared to the fBm power law, which was chosen to be similar to the power laws measured on Gaussian components of the multiscale segmentation performed by \citetads{2014MNRAS.440.2726R, 2019A&A...628A..33R}, the random cascade model power laws have a shallower power law than those measured for the non-Gaussian components. Because of the normalisation factor $C_\ell$, the power law of the random cascade models become independent of the original fBm.

\begin{table}
\centering
\small
\caption{Power laws fitted to the power spectra of Fig. \ref{fig:pow_spec_models}}
\label{tab:pow_spect_fits_models} 
\begin{tabular}{lc}
\hline\hline
& Power law $(\gamma)$ \\ \hline

fBm & $3.01\pm 0.01$ \\
efBm & $3.12\pm0.01$  \\
Isotropic random cascade & $1.87\pm0.01$  \\
Directional random cascade & $1.70\pm0.01$  \\

\hline
\end{tabular}
\end{table}

\section{Multifractal analysis}\label{sec:Multifractal}

The previous section described that the sequence of products by an independent random variable produces an increasingly non-Gaussian statistics towards smaller scales. In turbulence analysis, this effect is usually measured through structure functions of the centroid of the line-of-sight projected velocity increments, defined as

\begin{equation}
S_p(\ell)=\langle |\delta C_{\ell}(\mathbf{r})|^p\rangle \propto \ell^{\zeta_p},
\label{eq:struct_function}
\end{equation}

\noindent where $\delta C_{\ell}$ is the centroid velocity increment, $\delta C_{\ell}(\mathbf{r})=C(\mathbf{r})-C(\mathbf{r}+\ell)$, $p$ is the statistical moment of the distribution, and the average $\langle \rangle$ is calculated over all possible velocity increments separated by $\ell$. When $p$ increases, the velocity increment structure functions give a greater weight to rare events \citepads{2015MNRAS.446.3777B}. The \citetads{1941DoSSR..30..301K} turbulence theory predicts that the constant energy cascade in a turbulent medium yields an exponent $\zeta_p = p/3$ for every $p$. The constant energy cascade of \citetads{1941DoSSR..30..301K} for a subsonic non-compressive turbulent medium results in a superposition of random Gaussian fluctuations from large to small spatial scales, which is well represented by stochastic monofractals. This can be completely described statistically with a single power law. The multifractal nature of a medium induced by turbulence intermittency means that the scaling behaviour changes locally in the medium \citepads{1995turb.book.....F}. In the case of the log-normal theory of K62 \citepads{1962JFM....13...82K}, for example,  $\zeta_p$ becomes quadratic in $p$ such that $\zeta_p=\frac12 \mu_l p/3 (p/3-1),$ where $\mu_l$ quantifies intermittency. For a general description that is not only associated with turbulence, a multifractal scaling can be seen as a collection of interwoven fractal subsets with a range of scaling exponents $h$, and fractal (Hausdorff) dimension $D(h)$ \citepads{2001ApJ...551..712C}.

To estimate the multifractal spectrum $D(h)$, we chose the probabilistic reformulation of turbulence intermittency analysis in terms of wavelet transforms instead of increment structure functions \citepads{1993PhRvE..47..875M, 1995PhyA..213..232A}. The partition functions are here defined in terms of the wavelet coefficients:

\begin{equation}
Z_q(\ell) = \sum_{\vec{x}} \sum_{j=0}^{N_{\theta}-1}  |\tilde{f}(\ell,\vec{x},\theta_j)|^q.
\label{eq:partition_functions}
\end{equation}

\noindent Equation \ref{eq:partition_functions} can be considered as the equivalent of the structure function defined in Eq. \ref{eq:struct_function}, used here in models of column density maps, where $q$ is analogous to $p$ and gives a greater weight to rare events. This approach has been used to study the multifractal spectrum of HI maps by \citetads{2006ApJS..165..512K} through the formalism of the wavelet transform modulus maxima (WTMM) method \citepads{2000EPJB...15..567A}. The main difference with our analysis is that all wavelet coefficients are used to calculate the partitions functions, instead of only a subset of maximum values among the modulus of wavelet coefficients for the WTMM method. We also used the same wavelet as was used in the current work and in the MnGSeg technique \citepads{2019A&A...628A..33R}, the complex Morlet wavelet defined in Eq. \ref{eq:Morlet_Fourier}, instead of the derivative of a Gaussian, the DoG wavelet.

From these partition functions, we can define the scaling exponent $\tau(q)$ according to the power-law fit,

\begin{equation}
Z_q(\ell) \propto \ell^{\tau(q)}.
\label{eq:tauq}
\end{equation}

\noindent In equation \ref{eq:tauq}, $\tau(1)$ corresponds to the Hurst exponent $H$ that is related to the fractal dimension $d_f=E-H$, where $E$ is the Euclidean dimension of the image \citepads{1998A&A...336..697S, 2006ApJS..165..512K}. The index $\tau(2)$ is related to the second-order power law $\gamma$ measured by the Fourier power spectrum. The relation between the exponent $H$ and the power spectral index $\gamma$ is given by

\begin{equation}
\gamma = 2 + 2H \mbox{.}
\label{eq:Hvsgamma}
\end{equation}

\noindent Properly speaking, the Hurst exponent $H$ is generally used to characterise a monofractal image or surface, as is the case for fBms. In the case of a multifractal image, where the scaling exponent changes from point to point, this exponent becomes a local quantity and is generally referred to as the Hölder exponent $h(\vec{x})$. For a monofractal image, the function $\tau(q)$ is a linear function of $q$ of slope $H$. For a multifractal image, the function $\tau(q)$ is nonlinear and presents a collection of Hölder exponents, where $h=d\tau(q)/dq$.

As noted by \citetads{2006ApJS..165..512K}, two realisations of a given stochastic process are not a priori expected to have the same set of Hölder exponents. For this reason, multifractal analyses should be performed on several realisations of the same process. For this analysis, each model is realised on $1024\times1024$ images 64 times and averaged following the quenched averaging method \citep{PhysRevE.50.243}, where

\begin{equation}
e^{\langle \ln Z_q(\ell)\rangle} \sim \ell^{\tau(q)}, \qquad \ell \rightarrow 0^+.
\label{eq:quenched_averaging}
\end{equation}

\begin{figure*}
\centering
\includegraphics[width=0.75\textwidth]{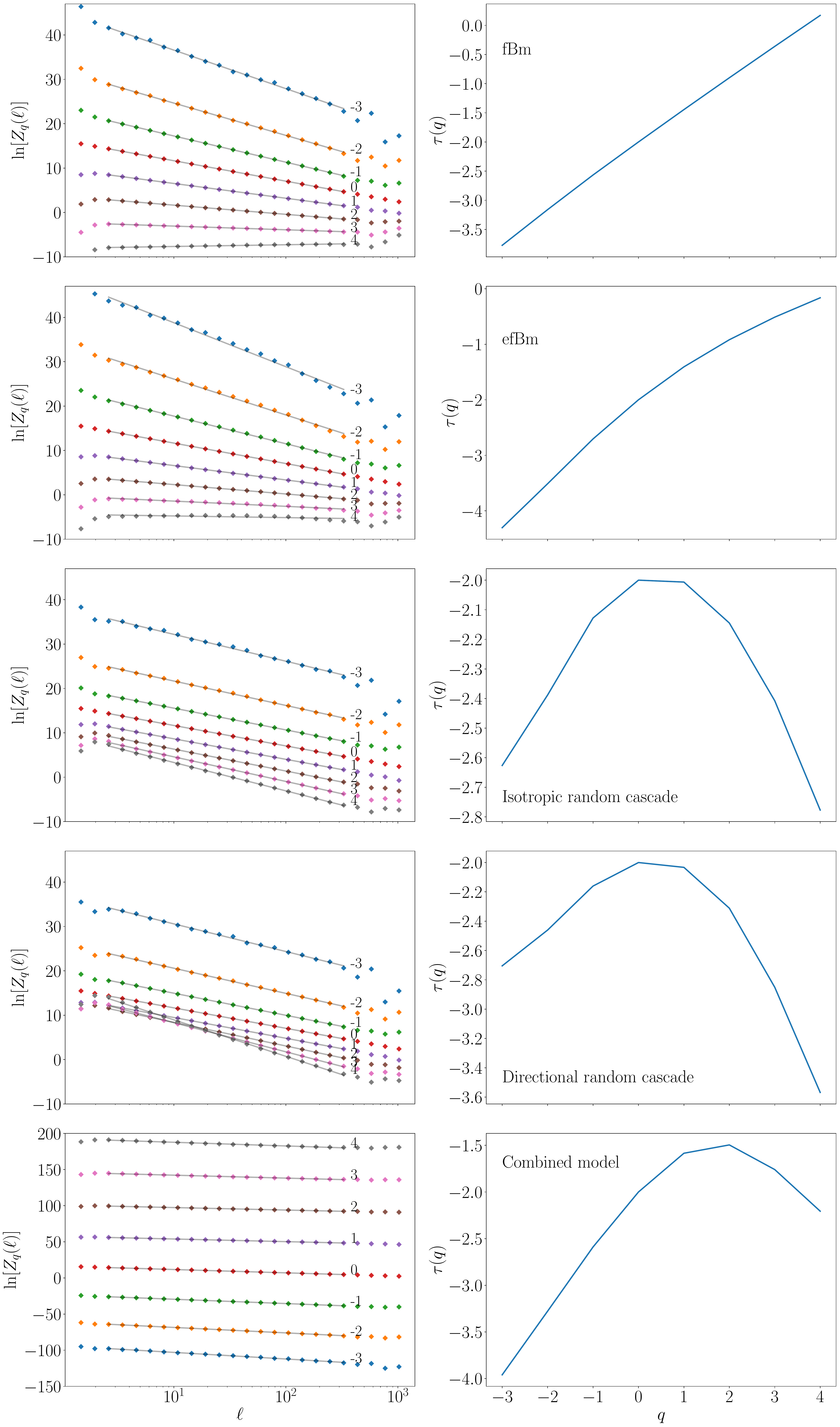}
\caption{Partition functions $Z_q(\ell)$ and the derived $\tau(q)$ functions according to their power-law fits for the quenched averaging of 64 realisations of the four models presented in Fig. \ref{fig:fractal_models} and the combined model presented in \S\ref{sec:hierarchy}.}.
\label{fig:tauq_figure}
\end{figure*}

\noindent The $Z_q(\ell)$ and $\tau(q)$ functions for the four models presented in this paper and the combined model presented in \S\ref{sec:hierarchy} are plotted in Fig. \ref{fig:tauq_figure}. As expected for the first fBm model, $\tau(q)$ is close to a linear function of $q$ of slope $H \simeq 0.55$, which corresponds to $\gamma \simeq 3.1$ (see Eq. \ref{eq:Hvsgamma}). On the other hand, the efBm model and the two random cascade models have a nonlinear $\tau(q)$ function, as expected for a multifractal geometry.

The multifractal spectrum $D(h)$, also called the singularity spectrum, is obtained from the Legendre transform of the partition function scaling exponent $\tau(q)$:

\begin{equation}
D(h) = \textrm{min}_q\left[qh-\tau(q)\right].
\label{eq:Legendre_transform}
\end{equation}

\noindent The function $D(h)$ measures the distribution of all the locations with a given scaling exponent $h(\mathbf{x}),$  that is, the Hausdorff dimension of that set of points. In other words, $D(h)$ shows the manner in which regions of different scaling exponents fill space \citepads{2001ApJ...551..712C}.

\begin{figure}
\centering
\includegraphics[width=0.48\textwidth]{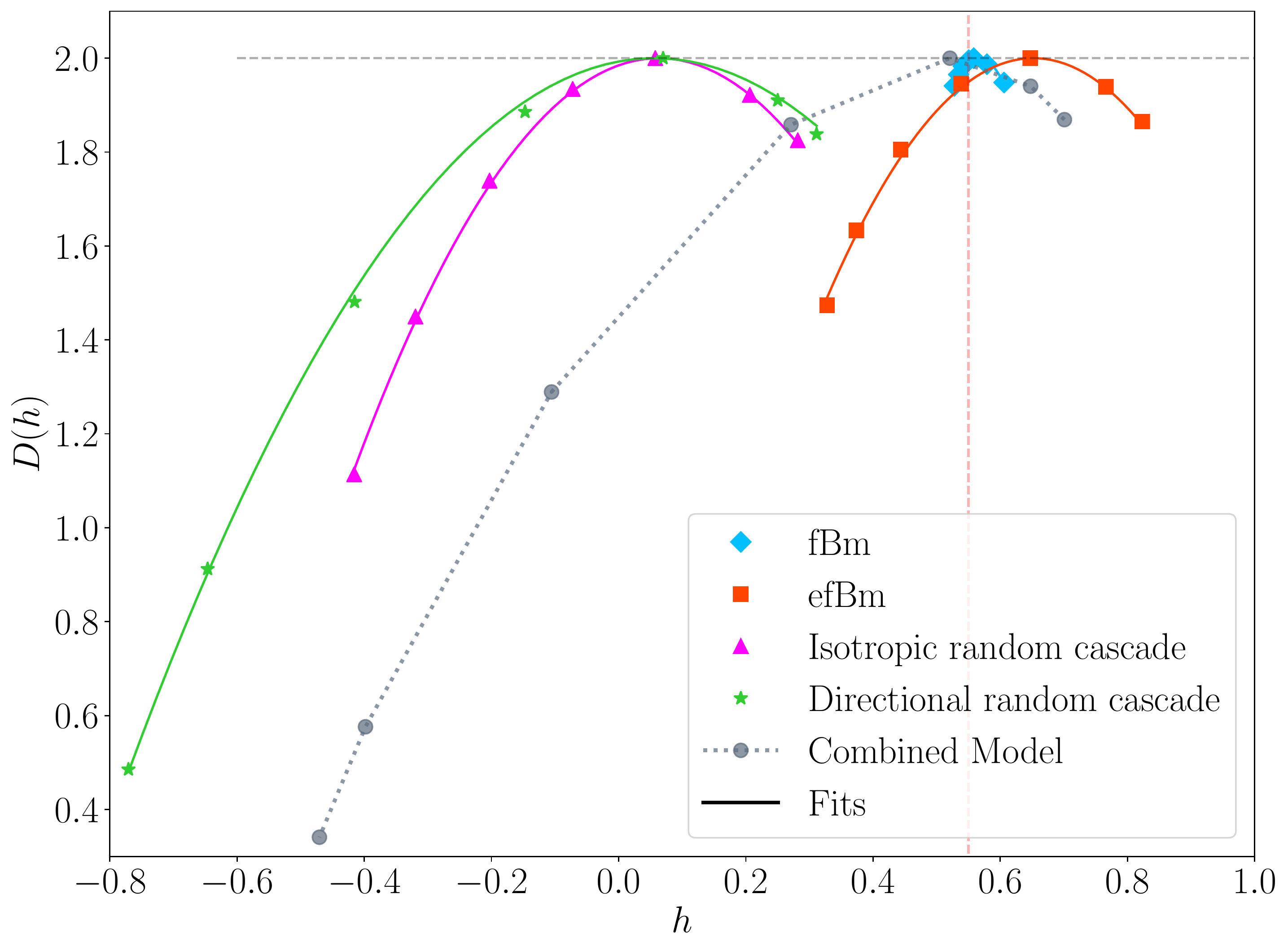}
\caption{Multifractal spectrum $D(h)$ derived from the $\tau(q)$ functions shown in Fig. \ref{fig:tauq_figure}. The dotted red line corresponds to $h=0.55$, the Hurst exponent imposed on the fBm model.}
\label{fig:Dh_figure}
\end{figure}

The resulting multifractal spectra for the models are shown in Fig. \ref{fig:Dh_figure}. The spectra were fitted with a parabola according to the so-called log-normal model \citepads{2010ApJ...717..995K, 1995turb.book.....F},

\begin{equation}
D(h) = E - \frac{(h-\mu_h)^2}{2\sigma_h},
\label{eq:fit_lognormal}
\end{equation}

\noindent where $E=2$, and $\mu_h$ and $\sigma_h$ are free parameters. The fitted values are listed in Table \ref{tab:D(h)}. The central value $\mu_h$ for the fBm corresponds well to the Hurst exponent $H = 0.55$ ($\gamma = 2+2H = 3.1$; the power-law index of the fBm) and has the smallest width $\sigma_h \simeq 0.01$ compared to the three other models, which have a $\sigma_h$ an order of magnitude higher. As expected, for a monofractal model, the distribution almost collapses to one point, but the multiplicative models of multifractal nature display a wide range of scaling exponents with a range of fractal dimensions. An independent multifractal analysis of the models has also been performed using the WTMM method \citepads{2000EPJB...15..567A, 2006ApJS..165..512K}. This second analysis gave the same results within the numerical uncertainties.

\begin{table}
\centering
\small
\caption{Parabola fitted value to $D(h)$ functions of Fig. \ref{fig:Dh_figure}}
\label{tab:D(h)} 
\begin{tabular}{lcc}
\hline\hline
& $\mu_h$ & $\sigma_h$  \\ \hline

fBm & $0.57\pm 0.01$ & $0.01\pm 0.01$\\
efBm & $0.65\pm0.01$ & $0.10\pm0.01$\\
Isotropic random cascade & $0.06\pm0.01$ & $0.13\pm0.01$  \\
Directional random cascade & $0.07\pm0.01$ & $0.22\pm0.01$  \\

\hline
\end{tabular}
\end{table}

According to the multifractal analysis, the fBm model represents well a non-intermittent turbulent medium with a constant mean energy dissipation rate on which all statistically averaged quantities depends. Consequently, because the fluctuating quantities modelled by an fBm model are  statistically well described by a single scaling exponent, $H$ or $\gamma$, its scaling index can also be estimated with its Fourier power spectrum. A turbulent field with an intermittent energy dissipation rate over the field affects other averaged quantities, such as the velocity or the density fluctuations, and they cannot be measured with a single Fourier power-spectrum index.  When the multifractal analysis is applied on the efBm, the isotropic and directional random cascade models demonstrate that these models can reproduce the intermittent behaviour of fully developed turbulence. However, the directional random cascade model is the only model of those presented here that is able to visually reproduce the filamentary aspect of the ISM.

The multifractal analysis we performed on our synthetic models has many differences compared to the recent study presented by \citetads{2018MNRAS.481..509E}, and we caution against comparing the two approaches. As pointed out by \citetads{2006ApJS..165..512K}, the box-counting approach used by \citetads{2018MNRAS.481..509E} and \citetads{2001ApJ...551..712C} is usually applied for the analysis of multifractal singular measures. The authors extrapolated this approach to continuous functions (i.e. 2D images) by considering $P_i$ as the sum of the pixel brightness for all pixels contained in the $i$th box. One consequence of this approach is the trivial estimate of the scaling exponent $h = 2.0$ for all fields instead of its relation with the Fourier spectrum power-law index, as demonstrated in our study. Alternatively, \citetads{1995PhyA..213..232A} suggested the use of wavelet functions for the multifractal analysis of continuous functions as a natural oscillatory variant and generalisation of the box function.

Furthermore, \citetads{2006ApJS..165..512K} showed that a multifractal analysis must be performed on several realisations of the same process with a well-defined averaging protocol. Our analysis was therefore realised on 64 images of similar synthetic models. Importantly, \citetads{2006ApJS..165..512K} also pointed out that the calculation of the $Z_q(\ell)$ function is increasingly less accurately estimated at large scales, especially for high values of $|q|$. For this reason, great care should be exercised when considering the range of $q$ values used in a multifractal analysis. The determination of this range of statistical order moments should not be arbitrary, but rather dictated by the availability of statistics in the analysed dataset to ensure statistical convergence. The range of $q$ values can be robustly and objectively determined (see \citetads{2006ApJS..165..512K} Eq. 32 and Figures 8 and 25). The multifractal characterisation of a medium or a surface is a very sensitive analysis, and we advise care in the choice of the approach according to the type of data and the size of the sample exhibiting a similar statistic. Unfortunately, several multifractal studies use $q$ values are high as $|q|=20$ without justification \citepads{2018MNRAS.481..509E, 2001ApJ...551..712C}, which can lead to numerical instabilities that may very well create artificial multifractal signatures.

\section{Hierarchical nature of filaments}\label{sec:hierarchy}

The Fourier power spectra for the four models presented in sections \ref{sec:fractal} and \ref{sec:random_cascade} are shown in Fig. \ref{fig:pow_spec_models}. All models follow a single scaling power law. The fBm and efBm models have a similar power law, which is the power law that was initially imposed on the complex noise in the Fourier space for the fBm. The two random cascade models have shallower power laws of $\sim1.8,$ indicating an enhancement of small-scale structures due to the multiplicative process, but also due to the normalisation factor $C_\ell$.

We recall that the models presented here are purely statistical and do not attempt to reproduce all the complex non-linear relations of physical processes that shape the ISM. However, without perfectly reproducing observational features, such as the curved shape of some filamentary structures, these models succeed in reproducing fundamental statistical properties of star-forming regions, such as the connection between the diffuse background density fluctuations, filamentary structures, and the multifractal nature of these regions.

 \citetads{2014MNRAS.440.2726R, 2019A&A...628A..33R} showed using the MnGSeg technique that the dense coherent and often filamentary structures of the ISM also possess a hierarchical nature. They also demonstrated for two different regions of the ISM that the power spectrum of filamentary structures dominates intermediate and small scales without producing any break in the total Fourier power spectrum of the region. Because these contrasted features are the result of progressive non-Gaussian contributions towards the small scales, no kink or break is visible in the Fourier power spectrum. These progressive non-Gaussian contributions may or may not change the total power law measured by the Fourier spectrum because second-order moments alone cannot fully describe the statistical properties of non-Gaussian fields. A similar progressive non-Gaussian contributions is responsible for the anomalous scaling of structure functions of centroid velocity increments for orders $p\geqslant4$. This statistical description of the filamentary dense coherent structures observed notably in the dust continuum images differs from the model suggested by \citetads{2019A&A...626A..76R}, where the synthetic filaments were modelled with a Gaussian or Plummer profile. This model adds a contribution in terms of power only to the spatial scale corresponding to the width of the filament profile. In this case, the power law measured by the Fourier power spectrum is only produced by the fluctuating background. The filament model proposed here contributes in terms of power to a broad range of spatial scales and has a power law that is only limited in the case of real observations by the telescope transfer function. Moreover, the multiplicative operation of the model imposes a spatial coherence through spatial scales to the filamentary structures. The directional modified random cascade model is the first statistical model that can produce filamentary structures that follow a global log-normal density distribution and has a hierarchical nature that is measured with its Fourier power spectrum.

\begin{figure*}
\centering
\includegraphics[width=1.0\textwidth]{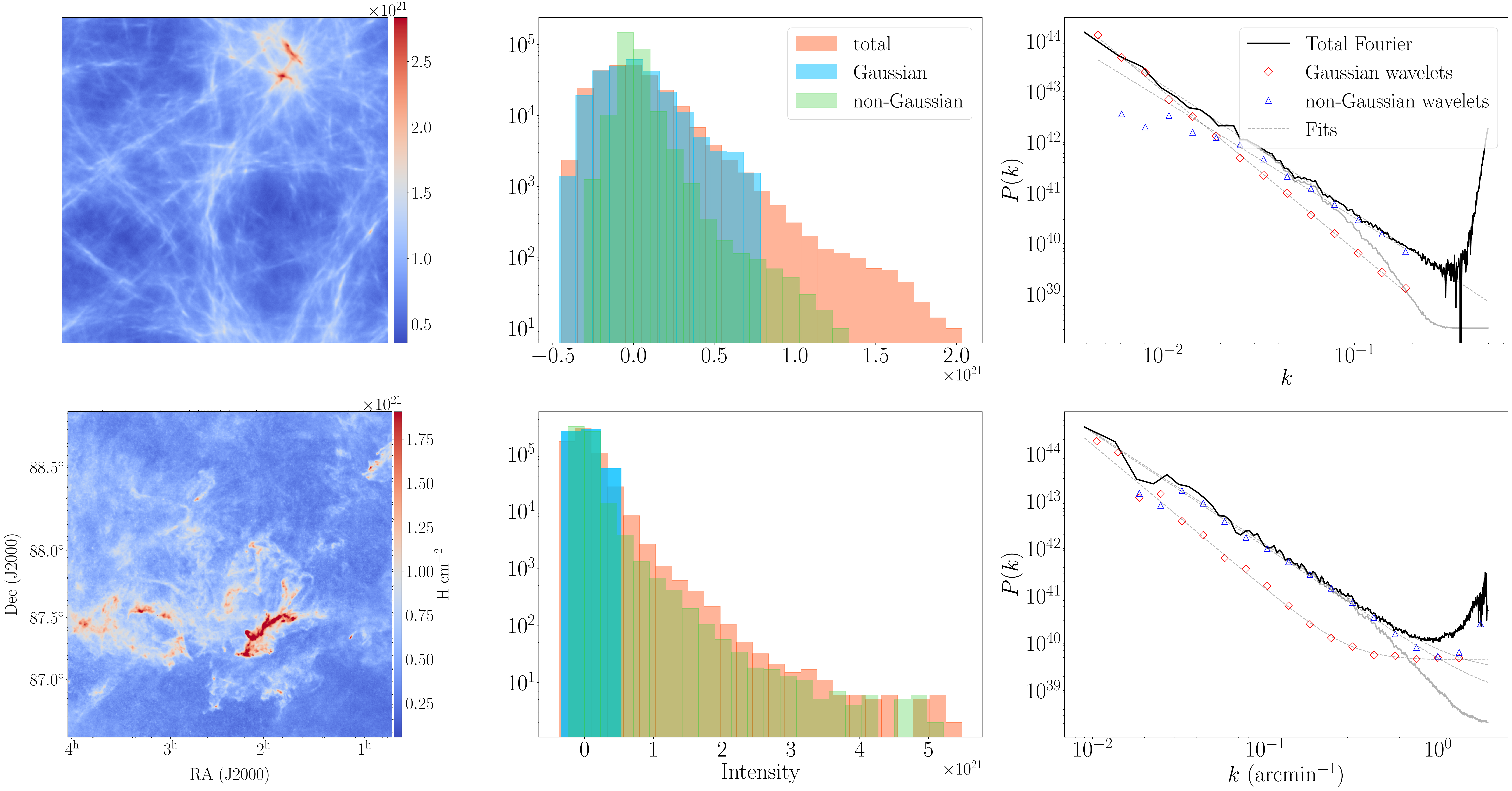}
\caption{Top left panel: Combined model, which is the addition of the fBm model in Fig. \ref{fig:fractal_models} (a) and the directional modified random cascade model in Fig. \ref{fig:fractal_models} (d) following Eq. \ref{eq:combined_model}. Top middle panel: Its segmented PDFs. Top right panel: Segmented power-spectra analysis. Bottom panels: Same figures for the Polaris flare region that was previously analysed by \citetads{2019A&A...628A..33R}.}
\label{fig:pow_spec_total}
\end{figure*}

\begin{table}
\centering
\small
\caption{Power laws for the power spectra of Fig. \ref{fig:pow_spec_total}}
\label{tab:pow_spect_fits_combined} 
\begin{tabular}{lcc}
\hline\hline
& Combined model & Polaris flare \\ \hline

Total Fourier & $2.53\pm 0.01$ & $2.38\pm0.02$ \\
Gaussian wavelet & $3.13\pm0.03$ & $3.05\pm0.07$ \\
Coherent wavelet & $2.33\pm0.05$ & $2.41\pm0.04$ \\

\hline
\end{tabular}
\end{table}

\begin{figure}
\centering
\includegraphics[width=0.48\textwidth]{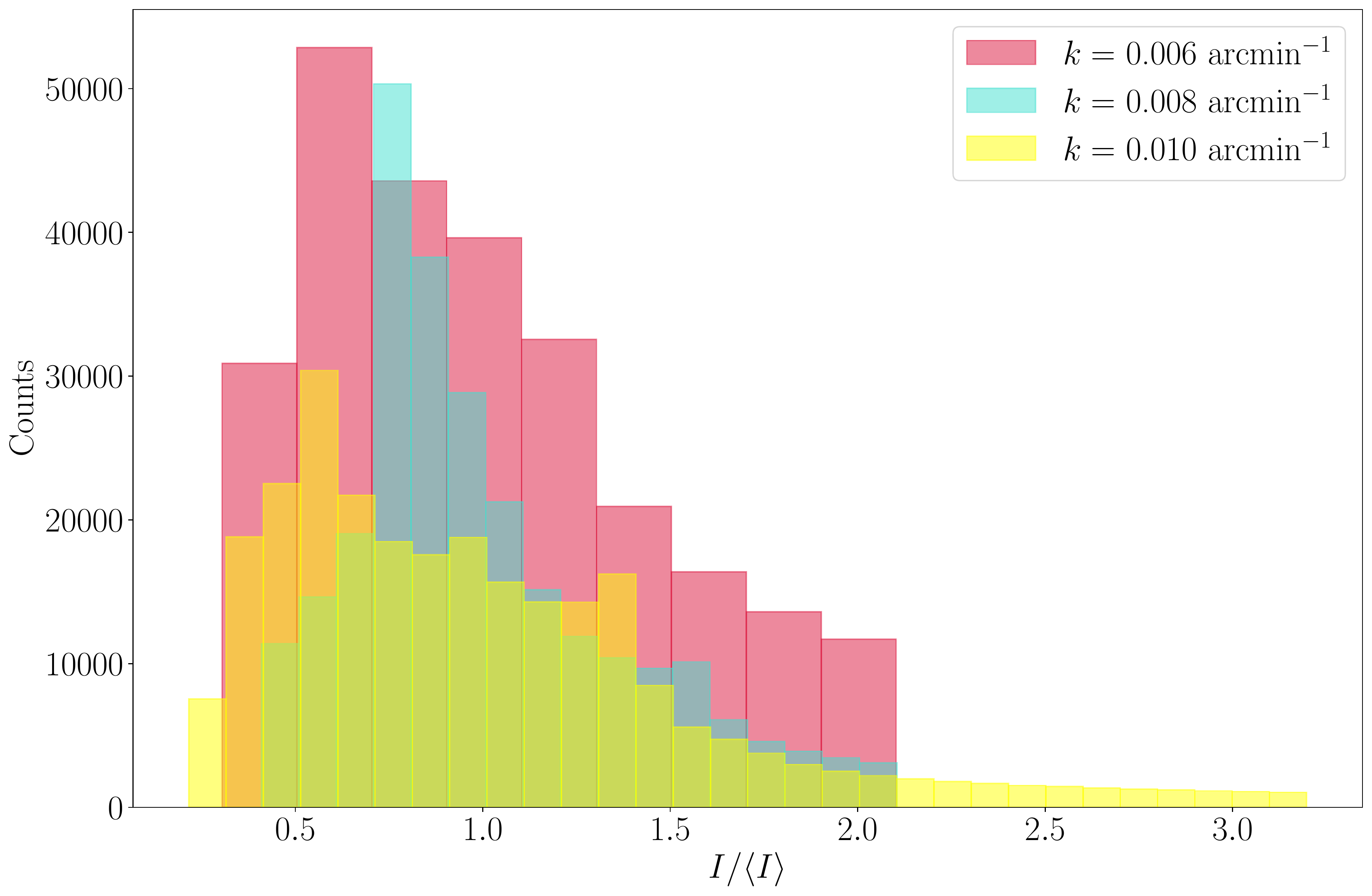}
\caption{PDFs as a function of scales for the combined model defined by equation \ref{eq:combined_model} that is shown in the top left panel of Fig. \ref{fig:pow_spec_total}.}
\label{fig:model_scale_PDFs}
\end{figure}

The coherent structures are formed through the product of a larger number of random phase linear waves at different spatial wavelengths. Dynamically, this effect might be associated with the collection of compressive processes that occur in the ISM. The final amplitude of the resulting filamentary structure depends on the phase coherence of the linear waves that model the shock-like wave fronts. The initial fBm image we used to construct the directional multiplicative random cascade is built with a random phase, which intends to be representative of the random mixing of the non-intermittent turbulence. As shown in Figures \ref{fig:fractal_models} (d), the hierarchy of filamentary
structures is formed through the random phase coherence of linear waves at different spatial wavelengths. We showed with the modified random cascade model that the product of initially random fluctuations is sufficient to reproduce some of the complex statistical properties of molecular clouds.

The fBm and the efBm models have a similar power law in Fig. \ref{fig:pow_spec_models}. This is due to the sensitivity limit of the Fourier power spectrum to non-Gaussian distributions as a function of scales. \citetads{2019A&A...628A..33R} proposed performing the segmentation of non-Gaussianities as a function of scales using complex wavelet transforms in order to detect the contribution of these structures and extract them. Previously, \citetads{2014MNRAS.440.2726R} showed that the efBm model failed to reproduce the shallower power law of non-Gaussianities as it is measured in observations. As shown in Fig. \ref{fig:pow_spec_models}, through the multiplicative process and the normalisation as a function of scales, the modified random cascade models succeed in producing a shallower power law than the Gaussian fBm by introducing increasingly more non-Gaussianities towards the small scales. Moreover, the distribution function and the Fourier power spectrum, when averaged over angles, are blind to the directional aspect of the modified random cascade model, which makes these measurements highly degenerate.

\section{Comparison with observations}

This study demonstrates that common statistical tools, such as the Fourier power spectrum and PDFs, present a high level of degeneracy. These techniques fail to capture the level of complexity of many hierarchical structures. These hierarchical properties observed in many star-forming regions can be represented by a collection of interwoven scaling exponents with different fractal dimensions. Unfortunately, because the multifractal analysis is high sensitive, the direct application of such methods on a limited number of observations that are affected by very similar physical processes is hazardous. However, knowing that these regions possess multifractal properties opens exciting new opportunities with regard to the development of new approaches to analysing their hierarchical structures and to understand how it affects the star formation activity of the regions. The development of such approaches is beyond the scope of this paper, but this section attempts to compare some fundamental properties observed in ISM structures with our models.

The top panels of Fig. \ref{fig:pow_spec_total} show the results of the addition of two models, the initial Gaussian fBm and the directional modified random cascade produced with the same fBm. The fBm power law was kept at 3.1, but the directional modified random cascade power law was artificially changed to 2.5 to match observations \citepads{2019A&A...628A..33R}. This combined model is defined by the following equation:

\begin{equation}
\rho_{\textrm{comb}}(\mathbf{x}) = \Gamma(\mathbf{x}) \otimes [A * \rho_{\textrm{fBm}}(\mathbf{x}) + \rho_{\textrm{dmc}}(\mathbf{x})] +N(\mathbf{x}),
\label{eq:combined_model}
\end{equation}


\noindent where $\Gamma(\mathbf{x})$ modelled the telescope transfer function with a Gaussian kernel with a full width at half maximum (FWHM) of 3.0 pixels. $\text{The crossed circle}$ is the convolution operation. $A$ is a constant, fixed to 5.0 in our model, to adjust the fBm power contribution compared to the multiplicative cascade. The standard deviation and the mean value of the combined model are also adjusted in order to compare the model with the Polaris flare region (lower panels of Fig. \ref{fig:pow_spec_total}). The noise level is modelled by $N(\mathbf{x})$ and also based on the noise level of the Polaris region. This combined model is an attempt to statistically reproduce the dual nature of molecular clouds described by \citetads{2004Ap&SS.292...89F}. In this description, star-forming clouds are seen as the combination of a diffuse component that dominates the large scales and has a fractal geometry, and a coherent component, which is well described by a network of filaments and dense cores. This description suggests a connection between the highly dynamical diffuse medium and dense molecular clouds. From a turbulence point of view, the large-scale fractal-dominated medium is connected to the non-intermittent and monofractal turbulence of \citetads{1941DoSSR..30..301K}, and the hierarchical coherent component is attributed to a multifractal subset of the field due to the intermittent behaviour of developed turbulence \cite[and references therein]{1995turb.book.....F}. We demonstrate using the multifractal extension of the MnGSeg technique \citepads{2019A&A...628A..33R} that the multiplicative models, including the efBm, are multifractal and thus have statistical properties similar to intermittent velocity fields of fully developed turbulence.

The top middle panel of Fig. \ref{fig:pow_spec_total} shows the distribution function of the combined model. The distribution function follows a log-normal distribution as for the efBm model. The segmented PDFs, according to the separation made by MnGSeg, correspond to the coloured distributions shown in Fig. \ref{fig:pow_spec_total}. The MnGSeg technique was applied on the wavelet coefficients in order to segment the non-Gaussianities as a function of the scale and the orientation of the Morlet wavelet transforms (see Appendix \ref{sec:Appen_MnGSeg} for more details). The Gaussian and the non-Gaussian (also called coherent) components of the map are reconstructed using relation \ref{eq:synthesis}, without adding the mean value $\mu_0$ to either of the two wavelet coefficient subsets, $\tilde{f}_{\rm Gaussian}(\mathbf{x},\ell,\theta_j)$ or $\tilde{f}_{\rm coherent}(\mathbf{x},\ell,\theta_j)$. The mean value $\mu_0$ was also subtracted from the total image PDF. Because the segmentation algorithm is applied in the wavelet space at multiple scales and for different orientations, the position of both components for the reconstructed maps is not exclusive. This property of the segmented map explains why their PDFs are overlapping. It is clear from the PDF analysis of the combined model (top middle panel of Fig. \ref{fig:pow_spec_total}) and for the Polaris flare molecular complex (bottom middle panel of Fig. \ref{fig:pow_spec_total}) that the exponential part of the PDFs is associated with the coherent structures in the maps. However, the coherent structures cover a wider range of intensity in the case of the Polaris region. This difference can be a sign that the multiplicative cascade alone cannot explain the highest density structures even for a low stellar activity region such as Polaris. This high density compared to the multiplicative cascade model might be explained by the effect of a global collapse and/or accretion mechanisms, which in addition to multiplicative processes have the ability to concentrate more the mass of the coherent structures \citepads{2019MNRAS.490.3061V}.

The right top panel of Fig. \ref{fig:pow_spec_total} presents the power spectrum analysis of the combined model. The Fourier analysis gives the total power spectrum of the model. The addition of the two models with two different power laws produces no clear break in the Fourier power spectrum, as for the Polaris flare region. The symbols are the averaged squared absolute value of complex wavelet coefficients at a certain spatial scale (see Appendix \ref{sec:Appen_MnGSeg}). The segmentation shows that the Fourier power spectrum is dominated by the filaments structures over a wide range of scales. The fitted power laws for the combined model and Polaris flare are shown in Table \ref{tab:pow_spect_fits_combined}.  \citetads{2019A&A...628A..33R} showed that the small-scale flattening in the Gaussian wavelet power spectrum was caused by the cosmic infrared background noise, which here was not modelled by the combined model.

As shown in the turbulence review by \citetads{1995turb.book.....F}, the models presented in Figures \ref{fig:fractal_models} and \ref{fig:pow_spec_total} satisfy the statistical spatial distribution of turbulent velocity fields and energy dissipation associated with non-intermittent monofractal media and/or intermittent multifractal media well. \citetads{2014MNRAS.440.2726R, 2019A&A...628A..33R} showed that these statistical properties, like the single power law of the Fourier power spectrum that is dominated by the contrasted coherent structures in field, are also characteristic of the gas column-density distribution observed through the infrared thermal dust emission. The direct dependences of the velocity and density fields in turbulence are difficult to determine. It is likely, however, that the multifractal nature of one may affect the statistical properties of the other. There is also a possibility that because of the non-linear nature and the hierarchical structures involved in this process, the mechanism of hierarchical collapse of molecular clouds such as described by \citetads{2019MNRAS.490.3061V} has the capacity of enhancing the multifractal nature of star-forming regions.

\section{Conclusion}\label{sec:conclusion}

We presented the first statistical model that can reproduce the fundamental gas distribution properties that have been seen in observations and numerical simulation of the ISM. It reproduces the ubiquitous log-normal distribution, the scale-free power spectrum, some aspects of the filamentary structures, and the multifractal geometry that is observed and measured in modern studies. The modified multiplicative random cascade model is easy to produce. The multifractal nature of star-forming regions has great implications on the hierarchal properties of their density distribution. The statistical modelling proposed in this paper indicates that dense coherent structures that are isolated through multiscale non-Gaussian segmentation exhibit a continuum of scaling exponents of different fractal dimensions. Multiplicative cascades can explain the origin of these structures. However, comparisons with observations seem to indicate that additional compressive effects, such as the gravitational collapse of the region, are necessary to explain the high density of observed star-forming regions compared to their diffuse monofractal background. The multifractal nature of these particular regions as a function of their derived properties or degree of evolution will be investigated in future works.

\section*{Acknowledgements}

This project has received funding from the European Union’s Horizon 2020 research and innovation programme under the Marie Sk\l odowska-Curie Grant Agreement No. 750920 and from the programme StarFormMapper under grant agreement No 687528. This work was also supported by the Programme National de Physique Stellaire and Physique et Chimie du Milieu Interstellaire (PNPS and PCMI) of CNRS/INSU (with INC/INP/IN2P3) co-funded by CEA and CNES. This research has made use of data from the Herschel Gould Belt survey (HGBS) project (http://gouldbelt-herschel.cea.fr). The HGBS is a Herschel Key Programme jointly carried out by SPIRE Specialist Astronomy Group 3 (SAG 3), scientists of several institutes in the PACS Consortium (CEA Saclay, INAF-IFSI Rome and INAF-Arcetri, KU Leuven, MPIA Heidelberg), and scientists of the Herschel Science Center (HSC). This research also made use of Astropy, a community-developed core Python package for Astronomy \citepads{2013A&A...558A..33A} and its affiliated package APLpy \citepads{2012ascl.soft08017R}.

\bibliographystyle{apj}
\bibliography{biblio}


\begin{appendix}

\section{non-Gaussian segmentation}\label{sec:Appen_MnGSeg}

The non-Gaussian segmentation of the wavelet coefficients is based on an iterative algorithm of denoising. The threshold $\Phi$ splits the non-Gaussian and the Gaussian wavelet coefficients at all scales and directions of the Morlet wavelet transform defined in Eq. \ref{eq:wavelet_transform}. The sequence defining $\Phi$ is

\begin{equation}
\left\{
                 \begin{array}{l}
           \Phi_{0}(\ell,\theta)=\infty \\
           \Phi_{n+1}(\ell,\theta)=q \sigma_{\ell,\theta}(\Phi_{n}(\ell,\theta)), \\
         \end{array}
         \right.
\label{eq:q_parameter}
\end{equation}

\noindent where $q$ (that is different from the $q$ in Eq. \ref{eq:partition_functions}) is a dimensionless constant that controls how restrictive the definition of non-Gaussianities is. The variance $\sigma^2_{\ell,\theta}$ is defined as

\begin{equation}
\sigma_{\ell,\theta}^2(\Phi)=\frac{1}{N_{\ell,\theta}(\Phi)}\sum_{\vec{x}} \mathbb{L}_{\Phi}( |\tilde{f}_{\ell,\theta}(\vec{x})|) |\tilde{f}_{\ell,\theta}(\vec{x})|^2,
\end{equation}

\noindent where

\begin{equation}
\mathbb{L}_{\Phi}(|\tilde{f}_{\ell,\theta}(\vec{x})|)=\left\{
                                                                                          \begin{array}{ll}
                                                                                  1 & \qquad \mathrm{if} \quad |\tilde{f}_{\ell,\theta}(\vec{x})| < \Phi \\
                                                                                  0 & \qquad \mathrm{else}. \\
                                                                                 \end{array}
                                                                                 \right.
\end{equation}

\noindent and

\begin{equation}
N_{\ell,\theta}(\Phi)=\sum_{\vec{x}} \mathbb{L}_{\Phi}( |\tilde{f}_{\ell,\theta}(\vec{x})|).
\end{equation}

\noindent The $q$ parameter varies as a function of $\theta$ and $\ell$. When the algorithm converges to an optimal value for the threshold $\Phi$, the skewness, the third moment of the distribution, of the Gaussian wavelet coefficient distribution is calculated. If the skewness is higher than a certain value, the parameter q is diminished by 0.1. This operation is repeated until convergence of the parameter q.

The segmented wavelet power spectra can be calculated by averaging the square absolute value of complex wavelet coefficients as a function of spatial scales,

\begin{equation}
P(\ell)=\frac{\delta \theta}{N_{\theta}N_{\vec{x}}} \sum_{\vec{x}} \sum_{j=0}^{N_{\theta}-1}  |\tilde{f}(\ell,\vec{x},\theta_j)|^2,
\label{eq:Fan_wavelet}
\end{equation}

\noindent where $\delta \theta=2\sqrt{-2\ln 0.75}/|\vec{k}_0|$, $N_{\theta}=\Delta \theta/\delta \theta$ is the number of directions $\theta$ needed to sample the Fourier space over the range $\Delta \theta,$ and $N_{\vec{x}}$ is the number of pixels in the image. The conversion between the spatial scale $\ell$ and the Fourier wavenumber $k$ is made through the relation $k=|\vec{k}_0|/l$. See \citetads{2019A&A...628A..33R} for more details.

\end{appendix}

\end{document}